\documentclass[aps]{revtex4}
\usepackage{epsfig}
\usepackage{amsmath,amssymb,color}

\parskip=\medskipamount

\newcommand{\eq}[1]{(\ref{#1})}
\newcommand{\fig}[1]{Fig.\ref{#1}}

\newcommand{\be}{\begin{equation}}
\newcommand{\ee}{\end{equation}}

\newcommand{\barr}{\begin{array}}
\newcommand{\earr}{\end{array}}

\newcommand{\beqn}{\begin{eqnarray}}
\newcommand{\eeqn}{\end{eqnarray}}

\newcommand{\bs}{\begin{subequations}}
\newcommand{\es}{\end{subequations}}

\newcommand\disp{\displaystyle}

\newcommand\scri{\scriptsize}

\newcommand{\la}{\left<}
\newcommand{\ra}{\right>}

\begin{document}

\title{Random Hierarchical Matrices: Spectral Properties and Relation to Polymers on
Disordered Trees}

\author{V.A. Avetisov$^1$, A.Kh. Bikulov$^1$, S.K. Nechaev$^2$\footnote{Also at:
P.N. Lebedev Physical Institute of the Russian Academy of Sciences, 119991, Moscow, Russia}}

\affiliation{$^1$N.N. Semenov Institute of Chemical Physics of the Russian Academy of
Sciences, 1199911, Moscow, Russia \\ $^2$LPTMS, Universit\'e Paris Sud, 91405 Orsay
Cedex, France}


\begin{abstract}

We study the statistical and dynamic properties of the systems characterized by an
ultrametric space of states and translationary non-invariant symmetric transition
matrices of the Parisi type subjected to "locally constant" randomization. Using the
explicit expression for eigenvalues of such matrices, we compute the spectral density for
the Gaussian distribution of matrix elements. We also compute the averaged "survival
probability" (SP) having sense of the probability to find a system in the initial state
by time $t$. Using the similarity between the averaged SP for locally constant randomized
Parisi matrices and the partition function of directed polymers on disordered trees, we
show that for times $t>t_{\rm cr}$ (where $t_{\rm cr}$ is some critical time) a
"lacunary" structure of the ultrametric space occurs with the probability $1-{\rm
const}/t$. This means that the escape from some bounded areas of the ultrametric space of
states is locked and the kinetics is confined in these areas for infinitely long time.

\end{abstract}

\maketitle

The explosion of the interest to the statistical properties of ensembles of random
matrices around 1950s has been motivated, in first turn, by a series of physical problems
in nuclear physics. Later the physical applications of the random matrix theory (RMT)
were extended to mesoscopics and statistical physics. One can mention recent works
\cite{johansson,spohn}, which deal with physical applications of the eigenvalue
statistics of ensembles of random matrices. The typical problems of RMT concern the
evaluation of the averaged spectral density, as well as distributions of "level spacings"
under the supposition that all matrix elements are independent randomly distributed
entries which take values in some specific ensemble \cite{mehta}.

Despite the standard RMT can be applied to a wide range of physical phenomena, RMT, in
its basic form, does not cover a particular class of {\em hierarchical} complex systems
important for glasses, real networks and proteins. The description of such complex
systems deals with the concept of hierarchical organization of energy landscapes
\cite{mez1,fra3}. A complex system is assumed to have a large number of metastable states
corresponding to local minima in the potential energy landscape. With respect to the
transition rates, the minima are suggested to be clustered in hierarchically nested
basins of minima, i.e. larger basins consist of smaller basins, each of these consists of
even smaller ones, {\it etc}. The basins of local energy minima are separated by a
hierarchically arranged set of barriers: large basins are separated by high barriers, and
smaller basins within each larger one are separated by lower barriers. Since the
transitions between the basins are determined by the passages over the highest barriers
separating them, the transitions between any two local minima obey the "strong triangle
inequality" for {\it ultrametric distances}. Hence, the hierarchy of basins possesses the
ultrametric geometry which has the natural visualization by the uniform $p$--adic (i.e.
$p+1$--branching) Cayley tree.

It might be said that the systems possessing the ultrametric geometry are hardly
compatible with the systems possessing the Euclidean geometry. Recently, however, the
interrelation between hyperbolic (i.e. tree--like) and Euclidean geometries has become a
topic of physics of disordered systems. It has been shown in \cite{fyod} that one can
construct random Gaussian translation—invariant potential landscape which reproduces all
the essential features of the Parisi landscapes directly in any finite--dimensional
Euclidian space. Also the question of isometric embedding of a uniform Cayley tree into
the 3D Euclidean space has been discussed in the work \cite{nech_voit}. The concept of
ultrametricity has been implemented in a number of toy models describing transport
phenomena and so-called "basin-to-basin" kinetics of disordered complex systems
\cite{ogi1,kohler1,becker1}. These models deal with various aspects of the {\it
ultrametric diffusion} -- a certain type of stochastic motion in a {\it nonrandom} energy
landscape with regular hierarchy encoded in the symmetric Parisi transition matrix,
$\bar{T}$. The model of ultrametric diffusion has been successfully applied to the
dynamics of proteins \cite{avetisov1}. The eigenvalues, $\{\bar{\lambda}_j\}$, of the
matrix $\bar{T}$ determine a hierarchy of relaxation times of the entire system and,
hence, define the kinetics constrained to an ultrametric landscape. The values of
$\bar{\lambda}_j$ are well known \cite{ogi1,avetisov1,vladimirov}:
\be
\bar{\lambda}_j=-p^{\gamma} \bar{T}^{(\gamma)} - (1-p^{-1})
\sum_{\gamma'=\gamma+1}^{\Gamma} p^{\gamma'} \bar{T}^{(\gamma')}
\label{eq:10}
\ee
were $\bar{\lambda}_{0}=0$ by definition, $\bar{T}^{(\gamma')}$ is the element  of the
Parisi matrix $\bar{T}$ related to the hierarchical level $\gamma$ of the $p$--adic tree,
and the summation runs up to the maximal level $\max[\gamma]= \Gamma$. Since $\bar{T}$ is
a kinetic matrix, the sum of all matrix elements in each column of $\bar{T}$ is equal to
zero.

The structural randomness is inherent for any disordered system. Thus, the representation
of the energy landscape by a {\it regular} hierarchy is rather ambiguous. In some cases,
like in spin--glass models, one supposes that the distribution of energy barriers is
fully random. However, for ultrametric energy landscapes it is more reasonable to suppose
that the barriers, belonging to the same hierarchical level, coincide only on the order
of magnitude. So, it is naturally to think that the transition rates between basins of
the {\it same} hierarchical level have random distribution around definite mean values
prescribed by the ultrametric hierarchy. Besides that, in real experiments we deal
usually not with individual molecule, but with an ensemble of molecules in the sample.
Therefore, the computation of an observable is connected with the averaging of a dynamics
in an individual ultrametric landscape over the ensemble of its realizations in the
sample.

\begin{figure}[ht]
\epsfig{file=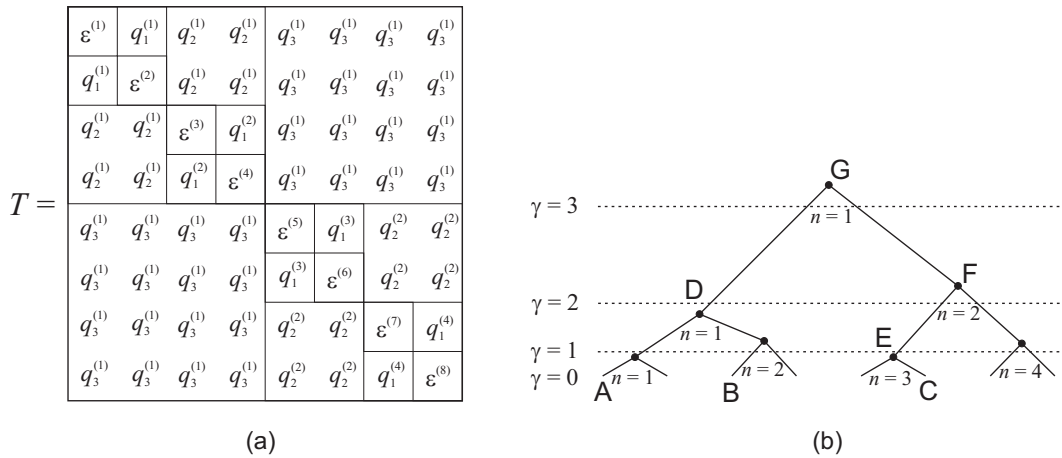,width=14cm} \caption{(a) Locally--constant randomized Parisi
matrix ($p=2$); (b) Locally--constant landscape corresponding to LCP matrix ($p=2$).}
\label{fig:1}
\end{figure}

To specify the model, consider two basins, belonging to the same hierarchal level. The
dynamic trajectories leading from one basin to the other one, pass through the common
"saddle point". The transition rates between the states in these basins depend only on
the location of the "saddle point", but not on the details of the behavior of the system
in the basins. This scenario implies that all elements of the Parisi--type matrix are
always equal inside a block, but they can fluctuate from one block to another one. Thus
we arrive at the randomized "locally--constant Parisi (LCP) matrix", $T$, it is depicted
in \fig{fig:1}a. Recall that $T$ is the kinetic matrix, hence the sum of matrix elements
in each column is zero.

Each matrix element, $T_{i,j}\equiv q_{\gamma}^{(n)}$, defines the probability to pass
over the corresponding barrier. Hence, the dimensionless barrier energy (the "barrier
height"), $u_{\gamma}^{(n)}=-\ln q_{\gamma}^{(n)}$ ($u_{\gamma}^{(n)}>0$) can be
introduced. As it has been pointed above, we suppose that
\be
u_{\gamma}^{(n)} = \left< u_{\gamma}^{(n)}\right> + \xi_{\gamma}^{(n)}
\label{eq:2_0}
\ee
where $\left<u_{\gamma}^{(n)}\right>$ is the mean barrier height in a given hierarchy
level, $\gamma$, and $\xi_{\gamma}^{(n)}$ describes the fluctuations around this mean
value. The mean values $\la u_{\gamma}^{(n)} \ra$ fix the structure of the nonrandom
(i.e. regular) Parisi matrix. If the barrier heights grow (in average) linearly with
$\gamma$, we set $e^{-\left<u_{\gamma}^{(n)}\right>}  = \la q_{\gamma}^{(n)} \ra =
p^{-(\alpha+1)\gamma}$, where $\alpha$ is a linear landscape "slope" ($\alpha\ge 0$)
\cite{ogi1,avetisov1}.

Thus the problem of stochastic motion in the randomized ultrametric landscapes has
appeared. The first important question concerns the computation of the spectral density
of the system described by the randomized Parisi--type symmetric matrix $T$ shown in
\fig{fig:1}a. Besides that, we analyze some probabilistic properties of the averaged
"survival probability" using the analogy with statistics of polymers on disordered trees.
For any distribution of $\xi_{\gamma}^{(n)}$ the spectral density of the matrix $T$ can
be, in principle, computed numerically. However since the locally--constant Parisi matrix
$T$ defines the new class of random matrices not yet considered in the literature, it is
worth understanding the analytic structure of spectral density of $T$ for the Gaussian
distribution of $\xi_{\gamma}^{(n)}$. In what follows we suppose that
$|\xi_{\gamma}^{(n)}|\ll 1$ in \eq{eq:2_0}. That allow us to truncate the power series
for $q_{\gamma}^{(n)}$ in $\xi_{\gamma}^{(n)}$ at the linear term:
\be
q_{\gamma}^{(n)} \approx e^{-\left<u_{\gamma}^{(n)}\right>}\left(1 - \xi_{\gamma}^{(n)} +
O\left( (\xi_{\gamma}^{(n)})^2\right)\right) \label{eq:2_1}
\ee
The non-negativity of the matrix elements of $T$ is ensured by the condition
$|\xi_{\gamma}^{(n)}|\ll 1$.

The randomized matrix $T$ can be viewed as an ultrametrically organized system of
barriers defined by a nonuniform $p$--adic tree -- see \fig{fig:1}a ($p=2$). Each matrix
element of $T$ represents a transition probability over the largest barrier separating
any two different states located on the lowest level $\gamma=0$. For example, the states
A and B are separated by the barrier in the branching point D, and the transition
probability from A to B is $q_2^{(1)}$; the states A and C are separated by the barrier
in the branching point G and the transition probability from A to C is $q_3^{(1)}$ (see
\fig{fig:1}a).

The eigenvalues of LCP matrix have been derived for the first time in \cite{avetisov2}
using the elements of the $p$--adic analysis \cite{vladimirov}. The construction of
$\lambda_{\gamma,n}$ has a very transparent geometric interpretation. Define a pair of
numbers $(\gamma,n)$, were $\gamma$ is the hierarchy level ($1\le \gamma \le
\Gamma=\max[\gamma]$) and $n$ enumerates the blocks in the same hierarchy level ($1\le n
\le p^{\Gamma-\gamma}$), as it is shown in \fig{fig:1}a. Then the eigenvalue
$\lambda_{\gamma,n}$ of LCP matrix for $p$--adic tree, reads:
\be
\lambda_{\gamma,n}=-p^{\gamma} q_{\gamma}^{(n)} - (1-p^{-1})
\underbrace{\sum_{\gamma'=\gamma+1}^{\Gamma} p^{\gamma'} q_{\gamma}^{(n')}}_{\Sigma}
\label{eq:1}
\ee
In Eq.\eq{eq:1} $q_{\gamma}^{(n)}$ are the elements of the matrix $T$ shown in
\fig{fig:1}a, while the summation $\Sigma$ (despite it resembles much \eq{eq:10})
contains some ambiguities. It is clear from \fig{fig:1}a that any pair $(\gamma,n)$ fixes
some vertex of the non-regular $p$--adic tree. The sum $\Sigma$ in \eq{eq:1},
contributing to $\lambda_{\gamma,n}$, runs from the point $(\gamma',n')$, which is the
upper nearest neighbor of the point $(\gamma,n)$ along a $p$--adic tree, towards the root
point $(\Gamma,1)$. For example, let us compute the eigenvalue $\lambda_{\gamma=1,n=3}$,
$p=2$ (see \fig{fig:1}b). The first term in \eq{eq:1} is just the weighted contribution
from the point E$(\gamma=1,n=3)$, while the sum $\Sigma$ in this case is the weighted sum
of two terms coming from the points F and G (see \fig{fig:1}a). Hence,
$\lambda_{1,3}=-2^1 q_1^{(3)}- (1-2^{-1}) \left[2^2 q_2^{(2)} + 2^3 q_3^{(1)} \right]$.

Let $P(\xi_{\gamma}^{(n)})$ be the distribution function of the fluctuations
$\xi_{\gamma}^{(n)}$ of matrix elements $q_{\gamma}^{(n)}$ (see \eq{eq:2_1}). The
spectral density, $\rho(\lambda)$ of the matrix $T$ can be computed in the standard way:
\be
\rho(\lambda)=\frac{1}{\cal N} \sum_{\{\gamma,n\}} \la \delta(\lambda -
\lambda_{\gamma,n})\ra
\label{eq:3}
\ee
where ${\cal N}=p^{\Gamma}$ is the total number of eigenvalues of the matrix $T$ and
$\left<...\right>$ means averaging with the distribution function
$P(\xi_{\gamma}^{(n)})$.

Suppose that the distribution of $\xi_{\gamma}^{(n)}$ is Gaussian and does not depend
neither on the hierarchy level, $\gamma$, nor on the index $n$ counting different blocks
in the same hierarchy level:
\be
P(\xi_{\gamma}^{(n)})=\frac{1}{\sqrt{2\pi \sigma^2}}
\exp\left(-\frac{(\xi_{\gamma}^{(n)})^2} {2 \sigma^2} \right)
\label{eq:2}
\ee
Speaking formally, since the fluctuations of $\xi_{\gamma}^{(n)}$ in \eq{eq:2} are not
restricted, the distribution $P(\xi_{\gamma}^{(n)})$ in Eq.\eq{eq:2} is inconsistent with
the condition $|\xi_{\gamma}^{(n)}|\ll 1$ introduced above. Recall that the last
condition ensures the matrix elements $q_{\gamma}^{(n)}$ in \eq{eq:2_1} to be
non-negative. However in what follows we suppose $\left<(\xi_{\gamma}^{(n)})^2\right>
=\sigma^2 \ll 1$, derive the spectral density and check our results for selfconsistency
by comparing analytic and numeric results. With \eq{eq:1}, \eq{eq:3} and \eq{eq:2} in
hands we can easily compute the spectral density, $\rho(\lambda)$. Proceed first with an
auxiliary computation of $Q(\lambda,\gamma)=\la\delta(\lambda-\lambda_{\gamma, n=1})\ra$
for $n=1$ and arbitrary $\gamma$ ($1\le \gamma \le \Gamma$), where the averaging is
carried out with $P(\xi_{\gamma}^{(n=1)})$. The straightforward computations give the
following result for $Q(\lambda,\gamma)$ at $\sigma\ll 1$:
\be
Q(\lambda,\gamma) = \int_{-\infty}^{\infty}\frac{dx}{2\pi}\, e^{i\lambda x} \la
e^{-i\lambda_{\gamma,n} x} \ra_{\left\{P(\xi_{\gamma}^{(n=1)}),...,
P(\xi_{\Gamma}^{(n=1)})\right\}}  \approx  \int_{-\infty}^{\infty} \frac{dx}{2\pi}
\exp\left\{i x \left[\lambda + v_{\gamma}(\Gamma)\right] -\frac{x^2}{2} \sigma^2
u_{\gamma}(\Gamma) \right\}
\label{eq:4}
\ee
where
\be
v_\gamma(\Gamma) = p^{-\alpha\gamma}+ (1-p^{-1})\sum_{\gamma'=\gamma+1}^{\Gamma}
p^{-\alpha \gamma'}; \qquad u_\gamma(\Gamma) = p^{-2\alpha\gamma}+
(1-p^{-1})^2\sum_{\gamma'=\gamma+1}^{\Gamma} p^{-2\alpha \gamma'}
\label{eq:vu}
\ee
The integral in \eq{eq:4} can be easily evaluated. Keeping only the leading (quadratic)
terms of expansion in $\sigma$, we get
\be
Q(\lambda,\gamma) = \frac{1}{\sqrt{2 \pi \sigma^2 u_\gamma(\Gamma)}}
\exp\left\{-\frac{\left[\lambda+v_{\gamma}(\Gamma)\right]^2} {\sigma^2
u_\gamma(\Gamma)}\right\}
\label{eq:5a}
\ee
The approximation \eq{eq:4} obtained for the expansion \eq{eq:2_1} is valid for
$\left<(\xi_{\gamma}^{(n)})^2\right> \ll 1$, i.e. for $\sigma^2 \ll 1$. Let us note that
it would be interesting to continue the expansions in \eq{eq:2_1} and in \eq{eq:4}
keeping explicitly the terms of order of $\sigma^4$. The computation of the spectral
density in this case involves the Airy functions. The discussion of this question will be
published in the extended version of our work.

\begin{figure}[ht]
\epsfig{file=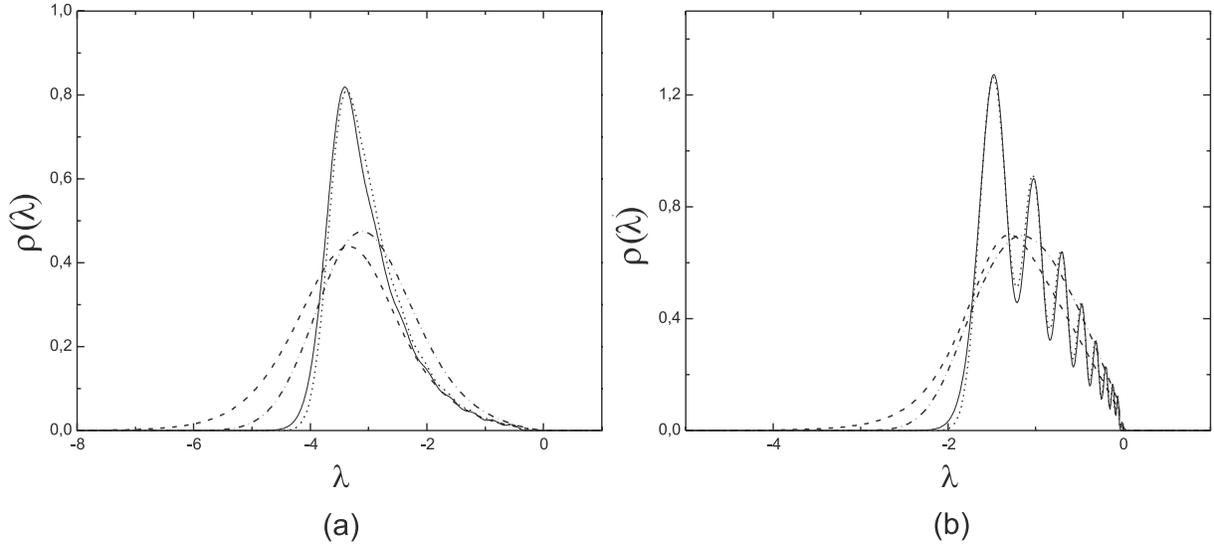,width=16cm} \caption{(Color online) Spectral density
$\rho(\lambda)$ for $\Gamma=8$: (a) $\alpha=0.1$: solid line -- explicit case and dotted
line -- linearized case for $\sigma=0.2$; dashed line -- explicit case and dotted--dashed
line -- linearized case for $\sigma=0.5$; (b) $\alpha=0.5$: notations the same as in
\fig{fig:3}a.}
\label{fig:3}
\end{figure}

Since the distribution function $P(\xi_{\gamma}^{(n)})$ does not depend on $n$, the
expressions \eq{eq:4}--\eq{eq:5a} hold for any of $1\le n \le p^{\Gamma-\gamma}$
eigenvalues on the hierarchical level $\gamma$. Introducing the number of eigenvalues on
the level $\gamma$ (i.e. the degeneracy), $g(\gamma)=p^{\Gamma-\gamma}$, we can rewrite
the spectral density, $\rho(\lambda)$, now as follows
\be
\rho(\lambda)=\frac{1}{p^{\Gamma}} \sum_{\gamma=1}^{\Gamma} g(\gamma) Q(\lambda,\gamma)
\label{eq:6}
\ee
We have computed numerically $\rho(\lambda)$ for the Gaussian distribution
$P(\xi_{\gamma}^{(n)})$, setting explicitly $q_{\gamma}^{(n)} = p^{-(\alpha +1)\gamma}\,
e^{-\xi_{\gamma}^{(n)}}$. Simultaneously we have considered the linearized case of matrix
elements, $q_{\gamma}^{(n)} \approx p^{-(\alpha +1)\gamma} (1-\xi_{\gamma}^{(n)})$. We
have compared in \fig{fig:3}a the explicit and linearized cases for $\Gamma=8$ and
$\alpha=0.1$. The same results for $\alpha=0.5$ are shown in \fig{fig:3}b. We see that
for $\sigma\lesssim 0.2$ the linearization of the matrix elements suggested in
\eq{eq:2_1} still makes sense for Gaussian distribution of energies. The analytic results
computed using \eq{eq:5a}--\eq{eq:6} coincide with the linearized case and are
indistinguishable in the figures.

The analysis of \eq{eq:6} for arbitrary $\alpha>0$ in the limit $\Gamma\to\infty$ allows
one to extract the asymptotic behavior of the tail of the spectral density
$\rho(\lambda)$ as $\lambda\to -\infty$:
\be
\rho(\lambda)\Big|_{\lambda\to -\infty} \approx \left\{
\begin{array}{rl} |\lambda|^{-\frac{\alpha-1}{\alpha}} e^{-\lambda^2} &
\mbox{for $\alpha \ge 1$} \medskip \\ |\lambda|^{-\frac{1-\alpha}{\alpha}} e^{-\lambda^2}
& \mbox{for $0<\alpha <1$} \end{array} \right.
\label{eq:7}
\ee

The information about the relaxation characteristics of any system (including the
hierarchical one) is encoded into the so-called "survival probability", $W(t)$, having a
sense of the probability to find a dynamical system in the initial state by the time $t$.
For a nonrandom kinetic Parisi matrix of $\Gamma$ hierarchy levels, the survival
probability is simply $\bar{W}(t,\Gamma)=(p-1) \sum\limits_{\gamma=1}^{\Gamma}
p^{-\gamma}\, e^{\lambda_{\gamma} t}+ p^{-\Gamma}$, where the eigenvalues,
$\lambda_{\gamma}$ are defined in \eq{eq:10} (recall that $\lambda_{\gamma}<0$ for
$\gamma=1,...,\Gamma$). For LCP transition matrices randomly distributed over the
ensemble, the survival probability is a random quantity. To make the survival probability
independent on the specific initial point on the tree, we consider $W(t,\Gamma)$ averaged
over all positions of initial states:
\be
\overline{W}(t,\Gamma) =(p-1)\sum_{\{\gamma,n\}} p^{-\gamma}\, e^{\lambda_{\gamma,n} t}
+p^{-\Gamma} = (p-1)\sum_{\gamma=1}^{\Gamma} p^{-\gamma} Z(t,\gamma,\Gamma) +p^{-\Gamma}
\label{eq:9}
\ee
where
\be
Z(t,\gamma,\Gamma)=\sum_{n=1}^{p^{\Gamma-\gamma}}  e^{\lambda_{\gamma,n} t}
\label{eq:11}
\ee
The sum in \eq{eq:11} runs over all $p^{\Gamma-\gamma}$ tree vertices of the level
$\gamma$ and the eigenvalues, $\lambda_{\gamma,n}$, are defined in \eq{eq:1}. Note that
despite $\overline{W}(t,\Gamma)$ does not depend on particular initial states on the
tree, it is still a random quantity since $Z(t,\gamma,\Gamma)$ and, hence,
$\overline{W}(t,\Gamma)$ depends (via Eq.\eq{eq:1}) on the ensemble of random barriers,
$\{\xi_{\gamma}^{(n)}\}$.

The direct computation of $\la W(t,\Gamma)\ra$ is a difficult problem, though some
important information about the probabilistic behavior of $\la W(t,\Gamma)\ra$ is
available by analyzing the distribution functions of the sub-sums $Z(t,\gamma,\Gamma)$.
Note that the eigenvalue, $\lambda_{\gamma,n}$, of the kinetic LCP matrix $T$ defines the
escape rate from the "basin" (i.e. sub-tree) with the root located in the point
$(\gamma,n)$ -- see \fig{fig:1}b. Hence, $Z(t,\gamma,\Gamma)$ is the partition function
of all escape rates at time $t$ from the hierarchy level $\gamma$.

Further consideration is based on the observation that $Z(t,\gamma,\Gamma)$ resembles the
partition function of the directed polymer on the disordered tree (DPDT) analyzed in
\cite{derrida}. Substituting the expansion \eq{eq:2_1} into \eq{eq:11}, and keeping only
the linear terms in $\xi_{\gamma}^{(n)}$ ($|\xi_{\gamma}^{(n)}|\ll 1$), we get
$Z(t,\gamma,\Gamma)= \tilde{Z}(t,\gamma,\Gamma) e^{-v_{\gamma}(\Gamma) t}$, where
$v_{\gamma}(\Gamma)$ is defined in \eq{eq:vu} and is independent on any particular path on
the tree, while $\tilde{Z}(t,\gamma,\Gamma)$ has the following expression:
\be
\disp \tilde{Z}(t,\gamma,\Gamma)=\sum_{\mbox{\scri all paths}} \exp\left\{t
\left(p^{-\alpha \gamma}\xi_{\gamma}^{(n)}+(1-p^{-1}) \sum_{\gamma'=\gamma+1}^{\Gamma}
p^{-\alpha \gamma'} \xi_{\gamma'}^{(n')}\right)\right\}
\label{eq:13}
\ee
The first summation in \eq{eq:13} is taken over all $p^{\Gamma-\gamma}$ vertices of
$\gamma$--level on the $p$--adic tree, and the sum in the exponent runs along the path on
the tree from the vertex $(\gamma,n)$ towards the root point (exactly as in \eq{eq:1}).

The function $\tilde{Z}(t,\gamma,\Gamma)$ satisfies the stochastic recursion (compare to
\cite{derrida})
\be
\tilde{Z}(\Gamma)=\exp\left[t(1-p^{-1})p^{-\nu\Gamma}\xi_{\Gamma}\right]
\sum_{j=1}^{p}\tilde{Z}_j(\Gamma-1)
\label{eq:14}
\ee
Introduce as in \cite{derrida} the averaged characteristic function, $G_{m}(x)$ by
\be
G_{m}(x) =\la \exp\left\{-\tilde{Z}(m)\,\exp\left[t(1-p^{-1})p^{-\alpha
m}x\right]\right\} \ra
\label{eq:15}
\ee
where $\gamma \le m \le \Gamma$. Using the factorization of $G_{m}(x)$ on the Cayley
tree, we come to the recursion for $G_{m}(x)$:
\be
G_{m-1}(x)=\int d\xi\, P(\xi)\, \left[G_{m}(p^{-\alpha}(x+\xi))\right]^p
\label{eq:16}
\ee
where $P(\xi)$ is the Gaussian distribution \eq{eq:2} for $\xi$. To equip \eq{eq:16} by
the initial condition, put in \eq{eq:13} $\gamma=\Gamma$ and get $\tilde{Z}(t,
\gamma=\Gamma,\Gamma)=e^{t p^{-\alpha \Gamma} \xi_{\Gamma}}$. Thus, we fix the initial
condition at the root of the tree by:
\be
G_{\Gamma}(x)=\la \exp\left\{-\exp\left[t(1-p^{-1}) p^{-\alpha
\Gamma}x\right]\,\exp\left[t p^{-\alpha\Gamma}\xi_{\Gamma}\right]\right\}
\ra_{P(\xi_{\Gamma})}
\label{eq:17}
\ee
The linear approximation in \eq{eq:13} implies the narrow distribution $P(\xi)$ with
$\sigma^2 \ll 1$. In this case the boundary condition \eq{eq:17} reads
\be
G_{\Gamma}(x)=\exp\left\{-\exp\left[t(1-p^{-1})p^{-\alpha \Gamma}x\right]\right\}
\label{eq:18}
\ee
Eqs. \eq{eq:16}, \eq{eq:18} set the problem. It is more convenient to revert the
direction along a tree by defining $n=\Gamma-m$ ($0\le n \le \Gamma-\gamma$). The case
$\alpha>0$ is a "contracting map" and will be analyzed elsewhere, while here we pay
attention to the case $\alpha=0$ (i.e. for $\la q_{\gamma}^{(n)} \ra = p^{-\gamma}$ ),
which is formally identical to the Derrida--Spohn model of DPDT \cite{derrida}. It is
known from \cite{derrida} that the solution of \eq{eq:16}, \eq{eq:18} in the continuum
approximation is related to the Kolmogorov--Petrovsky--Piscounov (KPP) equation
\cite{kpp} and has a travelling wave.

In terms of the work \cite{derrida}, for long distances on the Cayley tree, the partition
function, $G_n(x)$, is a travelling wave of the form $G_n(x)=w(x-fn)$, where the speed
$f$ is fixed by the initial condition, i.e. by the "inverse temperature", $\beta$, where
$\beta=t(1-p^{-1})$ (for $\alpha=0$). The speed, $f$, is the free energy of the system
per unit length in the long--length limit $f(t) = \frac{1}{\beta} \ln \left(p\int d\xi
P(\xi) e^{\beta \xi} \right)$. For $t<t_{\rm cr}$ the travelling wave propagates with the
speed $f(t)$, while for $t>t_{\rm cr}$ the speed $f$ is freezed at the critical value
$f(t_{\rm cr})$. The value $t_{\rm cr}$ is determined by the solution of the equation
$\left.\frac{d}{dt}f(t)\right|_{t=t_{\rm cr}}=0$. The overlap between two trajectories
starting from the common root point can be: i) 0 (with probability 1) for $t<t_{\rm cr}$,
and ii) either 0 (with probability $\pi(t)={\rm const}/t$) or 1 (with probability
$1-\pi(t)$) for $t>t_{\rm cr}$.

Despite the formal correspondence of our problem for $\alpha=0$ with DPDT is complete,
the interpretation of the behavior observed in \cite{derrida} in terms of our model is
not straightforward and deserves some discussion. The principal difference between the
partition function \eq{eq:13} and the one of DPDT consists in the following. Our kinetic
problem is defined on an $p$--adic tree, i.e. {\em only on the boundary} of an
ultrametric $p$--adic the Cayley tree and all kinetic properties of our model deal with
probabilities to cover some distance {\em along a tree's boundary} for given time
interval. To the contrary, the DPDT model is defined {\em in the "bulk" of the Cayley
tree}, i.e. on the whole set of tree's generations, $m$, and the travelling wave
propagates {\em along} $m$, from the root point, $m=0$, "downwards".

The overlap of the trajectories computed in \cite{derrida} allows one to connect the
"bulk" and the "boundary" behaviors. Since on small time scales $t<t_{\rm cr}$ (i.e. for
"high temperatures") the overlap of any two trajectories (starting from the root of the
tree) is zero, the boundary of the Cayley tree is uniform since there is no clustering.
Hence, the escape probability from any basin at the boundary of the Cayley tree, is
independent on the specific point on this boundary and the kinetics happens as for
nonrandom Parisi matrix. To the contrary, on large time scales $t>t_{\rm cr}$ (i.e. for
"low temperatures") the overlap of the trajectories signals the "probabilistic
non-uniformity" of the tree boundary: with the probability $\pi(t)$ the boundary is still
uniform, while with the probability $1-\pi(t)$ the boundary has a "lacunary" structure
such that the escape from some boundary basins is locked and the kinetics is confined in
these basin for infinitely long time.

The linear approximation \eq{eq:2_1} applied to \eq{eq:13} deserves some special care.
Actually, due to \eq{eq:2_1} the unphysical positive tail in the spectral density
$\rho({\lambda})$ could appear. Since the survival probability $\overline{W}(t,\Gamma)$
depends on all eigenvalues (see Eq.\eq{eq:9}), for sufficiently large $t$ these
unphysical tail could have a non-negligible effect in the observed behavior. In order to
eliminate such an influence, let us note that as $\sigma^2$ decreases, the linearized
behavior becomes better (see \fig{fig:3}). So, we fix $\sigma$ and define the maximal
time, $t_{\rm max}(\sigma)$, until which our predictions are valid and the interference
of unphysical tail is negligible. The estimation of $t_{\rm max}(\sigma)$ is very
straightforward. Namely, it follows from \eq{eq:9} that the major contribution to
$\overline{W}(t,\Gamma)$ at large $t$ comes from the relaxation modes, for which $t\,
\lambda_{\gamma}^{(n)} \sim 1$. At the same time one sees from \eq{eq:5a} that the
unphysical tail corresponding to $\lambda>0$ is exponentially small for $v_{\gamma} \gg
\sigma \sqrt{u_{\gamma}}$. For practical purposes we may use the estimate $v_{\gamma}
\gtrsim 2 \sigma \sqrt{u_{\gamma}}$. Using \eq{eq:vu} we get $\sigma \lesssim 0.5$. Thus,
for $\sigma \lesssim 0.5$ the influence of unphysical tail is exponentially suppressed
for any $t$.

In this letter we have analyzed the simplest statistical properties of randomized
Parisi--type kinetic matrices. We also believe that the observed relation between
dynamics on ultrametric landscapes described by LCP matrices and directed polymers on
disordered trees is very promising for establishing a connection between the elements of
$p$--adic analysis and disordered models on tree--like structures. Our work should be
considered a preliminary step in this direction. The straightforward generalization of
the obtained results concern an explicit account for the terms of order of
$|\xi_{\gamma}^{(n)}|^2$ in \eq{eq:2_1} for the spectral density and for the survival
probability.

We are grateful to K. Bashevoy and O. Vasilyev for illuminating numerical simulations, to
O. Bohigas and Y. Fyodorov for helpful discussions, and to M. Mezard for paying our
attention to the relation between LCP systems and DPDT. This work is partially supported
by the RFBR grant No. 07-02-00612a.

\end{document}